\documentstyle[prl,aps,floats,epsf]{revtex}

\begin{document}
\draft

\twocolumn[\hsize\textwidth\columnwidth\hsize\csname
@twocolumnfalse\endcsname
\title{ Rigidity of the elastic domain structure
near the boundary of its existence 
 in thin epitaxial films 
}
\author{A.M. Bratkovsky$^{1}$ and A.P. Levanyuk$^{1,2}$}
\address{$^{1}$Hewlett-Packard Laboratories, 1501 Page Mill Road, Palo Alto,
California 94304\\
$^{2}$Departamento de F\'{i}sica de la Materia Condensada, CIII, Universidad
Aut\'{o}noma de Madrid, 28049 Madrid, Spain}
\date{February 22, 2001}
\maketitle

\begin{abstract}

We consider an interesting and practically important case of elastic 
domain structure, which is the analogue of c/a domain pattern with 
90$^\circ$ walls
in perovskites, and is solvable analytically for arbitrary misfit strain.
There is no critical thickness, below which the domain structure
cannot exist,  when the ``extrinsic" misfit is zero and the domains
are of equal width.
At the boundary of polydomain-monodomain transition the period of 
the pattern diverges,  as does the dynamic  stiffness
of the domain  structure.  It is unlikely, therefore,  that one can
achieve a softness of the dielectric response of the c/a elastic domains  
in ferroelectric-ferroelastic thin films.

\pacs{77.80.Dj, 77.55.+f, 81.30.Dz}

\end{abstract}
\vskip 2pc ]
\narrowtext
Equilibrium domain structures in epitaxial ferroelectric films may appear
even in the case of complete compensation of the depolarizing electric
fields by the electrodes or finite conductivity of the film. This takes
place if the film also behaves as a ferroelastic. Formation of a
ferroelastic domain structure was first considered a while ago as a
mechanism to relax the misfit imposed by a substrate \cite{Roitburd76}. In
widely studied perovskite ferroelectrics, which are improper ferroelastics,
several elastic domain (and heterophase) structures were predicted \cite
{RoytburdYu,Pompe,Koukhar00}. One of the typical structures, so-called $%
c/a/c/a...$ domain pattern, is of a special fundamental and practical
interest. Firstly, the ratio of the widths of the $c$- and $a$-domains
changes with external conditions (film thickness and misfit strain). At some
parameters the domain pattern does not exist, i.e. there is a phase
transition between multi- and monodomain states. Although the corresponding
phase diagrams have been discussed by several authors (see \cite
{Pompe,Koukhar00} and references therein), the behavior close to the phase
transition was not studied. The model calculations\cite{Pertsev96} have
predicted the response of the $c/a$ domain structure to be large, even giant
for some specific values of the (``extrinsic'', see below) misfit $w$ and
the film thickness $l$, and some data was even interpreted in this way\cite
{erbil96}. The enhancement of the response was expected near the border of
existence of the $c/a$ structure in the $w-l$ plane, i.e. near the phase
transition. Previous analyses, mainly numerical, have not actually specified
the system parameters needed to realize the giant susceptibility. We
attempt, therefore, an exact solution of a simplest relevant model,
analogous to the actual $c/a$ structures. The present results put strict
constraints on the system parameters where the softness of the dielectric
response might be expected. It becomes clear that it would be very difficult
to realize the domain structure with such a property. As a corollary, the
existing data \cite{erbil96} cannot be interpreted as being due to a large
contribution of the elastic domain walls. We reveal in the present paper
some unusual features of the polydomain-monodomain phase transition and show
that the response of the system becomes more ''rigid'' in the vicinity of
this transition.

We present here the first analytical calculations of the energy of a domain
pattern analogous to the standard $c/a$ structure. The structure can be
viewed, without loss of generality, as a result of a ferroelastic transition
in an epitaxial film, which breaks the symmetry of the parent phase and the
substrate. All the strains are considered in the reference frame of the high
symmetry phase, and the $z$-axis is selected to be perpendicular to the
plane of the film. We consider the system far from the transition, where the
pattern consists of the domains having the spontaneous strain, $%
u_{xx}^{0}-u_{zz}^{0},$ of opposite signs, separated by the domain walls
inclined at 45$^{\circ }$ with respect to the $z$-axis, as dictated by the
elastic compatibility conditions and the lattice symmetry, Fig.~1. In
ferroelectric perovskites this would correspond to the 90$^{\circ }$ walls
separating the domains with the polarization parallel and perpendicular to
the plane. The film consists of $a-$ and $c-$type domains with the widths $%
a_{1}$ and $a_{2},$ respectively. From the exact expressions for the free
energy $F$ [see Eqs.(\ref{eq:Fh}),(\ref{eq:Fstray})], we calculate the
period $2a=a_{1}+a_{2}$ of the structure, the parameter of asymmetry $\delta
=$ $\left( a_{1}-a_{2}\right) /2a,$ and the corresponding stiffness $\kappa $%
. The stiffness $\kappa $ is closely related to the ``mechanical force
constant'' discussed in Ref. \cite{Pertsev96}. We shall consider {\em two}
stiffnesses: the usually measured {\em dynamic} stiffness $\kappa _{\infty }$%
, calculated at fixed period of the structure, and the {\em static}
stiffness $\kappa _{0},$ corresponding to the situation where the period of
domain structure is allowed to relax via creation or annihilation of the
domain walls. The latter corresponds to {\em very} slow processes, making
its observation very challenging. One certainly deals with the dynamic
stiffness $\kappa _{\infty }$ in applications. We shall show that the two
stiffnesses exhibit quite different behavior near the transition: the static
one diminishes, while the dynamic one diverges. We also study the change of
the stiffness with the film thickness and discuss the possibilities of
obtaining the films with small dynamic stiffness. %
%
\begin{figure}[t]
\epsfxsize=2.8in \epsffile{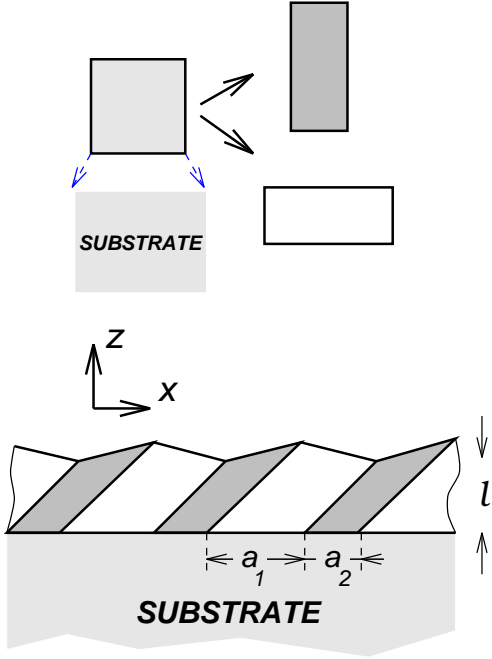}
\caption{Schematic of the ferroelastic phase transformation in epitaxial
film, when the ``extrinsic'' misfit between the parent phase and the
substrate is also present (top panel). As a result, the domain structure
with the period $2a=a_{1}+a_{2}$ is formed in the film of the thickness $l$
(bottom panel). }
\label{fig:fig1}
\end{figure}

We consider first the case where the sole origin of the misfit is the
spontaneous strain, $u_{xx}^{0}-u_{zz}^{0},$ appearing at the phase
transition. We shall call this misfit ``intrinsic'', to distinguish it from
``extrinsic'' misfit of any other origin. The latter includes e.g. the
misfit of parent phase with the substrate, and other, ``noncritical'',
strain components, which might appear during the phase transition. The
effect of extrinsic strain is a focus of the present paper. Similar to
previous authors \cite{Pertsev96} we neglect the uncompensated electric
fields, which would only increase the stiffness.

The method of calculating the energy of the domain structure is the same as
in our previous paper \cite{BLxy} but with different equations of state for
the stress tensor components $\sigma _{ik}$: 
\begin{eqnarray}
\sigma _{zz} &=&\left( \lambda +2\mu \right) e_{zz}+\lambda \left(
e_{xx}+e_{yy}\right) ,  \label{eq:szz} \\
\sigma _{xz} &=&2\mu e_{xz},  \label{eq:sxz}
\end{eqnarray}
with $\sigma _{xx}$ and $\sigma _{yy}$ components obtained from (\ref{eq:szz}%
) by cyclic permutation of $x,y,z$. Here $\lambda ,\mu $ are the Lam\'{e}
coefficients, and $e_{ik}=u_{ik}-u_{ik}^{0}$ is the elastic strain, with $%
u_{ik}^{0}$ the components of the spontaneous strain. In contact with
substrate, the film with symmetry-breaking misfit with the substrate must
split into domains in such a way that the average in-plane strain will be
zero, since the homogeneous strain in the substrate would cost an infinite
energy. {\em Without} the ``extrinsic'' strain the spontaneous strain would
alternate from domain to domain (which all would have the same width) as $%
u_{xx}^{0}=-u_{zz}^{0}\equiv \pm u_{0},$ $u_{yy}^{0}=0.$ On the other hand, 
{\em with} the ``extrinsic'' misfit strain $w$ the distribution of the
spontaneous strain in the domains would be 
\begin{equation}
u_{xx}^{0}(u_{zz}^{0})\equiv \pm (\mp )u_{0}+w,\qquad u_{yy}^{0}=w.
\label{eq:u0}
\end{equation}
We assume for the substrate the same equations of state, as (\ref{eq:szz}),(%
\ref{eq:sxz}), but with $u_{xx}^{0}=u_{yy}^{0}=u_{zz}^{0}=0.$

To find the energy of the domain structure, we have to determine the energy
of the homogeneous and inhomogeneous (stray) stresses. The energy of the 
{\em homogeneous} stresses $\left( F_{h}\right) ,$ which appears for domains
with non-equal widths (i.e. for $\delta \neq 0),$ can be easily found for an
epitaxial film with {\em any} domain structure, provided that the elastic
moduli in all domains are the {\em same}, and the non-linear effects can be
neglected. We can readily find $F_{h}$ from the general expression for the
elastic energy $F_{el}=-\frac{1}{2}\int dV\sigma _{ik}u_{ik}^{0}$ \cite
{BLxy,Mura} as 
\begin{eqnarray}
\frac{F_{h}}{{\cal A}} &=&-\frac{l}{2}(\bar{\sigma}_{xx}\bar{u}_{xx}^{0}+%
\bar{\sigma}_{yy}\bar{u}_{yy}^{0}+\bar{\sigma}_{zz}\bar{u}_{zz}^{0}) 
\nonumber \\
&=&Ml\left[ \left( \bar{u}_{xx}^{0}\right) ^{2}+\left( \bar{u}%
_{yy}^{0}\right) ^{2}+2\nu \bar{u}_{xx}^{0}\bar{u}_{yy}^{0}\right]
\label{eq:Fhgen} \\
&=&Mu_{0}^{2}l\left[ (\delta -\delta _{0})^{2}+w^{2}(1-\nu ^{2})/u_{0}^{2}%
\right]  \label{eq:Fh}
\end{eqnarray}
where $\delta _{0}=-w(1+\nu )/u_{0}$ is the relative extrinsic misfit, $\nu
=\lambda /2(\lambda +\mu )$ the Poisson's ratio, and $M=2\mu \left( \lambda
+\mu \right) /\left( \lambda +2\mu \right) \equiv E/2(1-\nu ^{2})$, where $E$
is the Young's modulus. The overbar here and below marks the averaging over the
film (domain pattern). The condition $\bar{\sigma}_{zz}=0$ was used together
with Eq.(\ref{eq:szz}) to obtain for the planar stresses $\bar{\sigma}%
_{xx}=2M(\bar{e}_{xx}+\nu \bar{e}_{yy}),$ $\bar{\sigma}_{yy}=\bar{\sigma}%
_{xx}(x\leftrightarrow y),$ where $\bar{e}_{xx}\equiv \bar{u}_{xx}-\bar{u}%
_{xx}^{0}=-\bar{u}_{xx}^{0}=-u_{0}\delta -w,$ $\bar{e}_{yy}=-\bar{u}%
_{yy}^{0}=-w,$ Eq.(\ref{eq:u0}). We have also used the absence of the
in-plane strains, $\bar{u}_{xx}=\bar{u}_{yy}=0$, imposed by the substrate,
and an obvious relation $\bar{u}_{xx}^{0}=\frac{1}{2}(1+\delta )(u_{0}+w)+%
\frac{1}{2}(1-\delta )(-u_{0}+w)=u_{0}\delta +w.$ Note that Eq.(\ref
{eq:Fhgen}) is rather general and can be applied to other kinds of domain
patterns in the epitaxial films (cf. \cite{BLxy,RoytburdYu}).

The {\em stray} energy of inhomogeneous stresses in a stripe-like domain
structure periodic in $x-$direction, Fig. 1, is to be found from exact
solutions for the strain field produced by the domains. The pattern is
defined by the distribution of spontaneous strains $u_{xx}^{0}\left(
x\right) $ and $u_{zz}^{0}\left( x\right) .$ The condition of the local
equilibrium, $\partial \sigma _{ik}/\partial x_{k}=0,$ gives two sets of two
equations for the displacement components $u_{x},$ $u_{z}$ in the film and
the substrate with the use of a standard relation $2u_{ik}=\partial
u_{i}/\partial x_{k}+\partial u_{k}/\partial x_{i}$. The boundary conditions
are given by the absence of stresses, $\sigma _{zz}=$ $\sigma _{xz}=0$ at
the free surface of the film ($z=l$) and at $z\rightarrow -\infty $. Both
the strain tensor and the displacement vector are continuous at the
film-substrate interface. The original system of partial differential
equations is reduced to a system of the ordinary differential
equations with the 
use of the Fourier transform and then solved, as described in detail in
Refs. \cite{BLx2stab,BLxy}, and we obtain the following simple expression
for the stray energy per area ${\cal A}:$

\begin{equation}
\frac{F_{stray}}{{\cal A}}=\frac{2Mu_{0}^{2}a}{\pi ^{3}}\sum_{n=1}^{\infty }%
\left[ 1-\left( -1\right) ^{n}\cos \pi n\delta \right] \frac{1-\left(
1+2k^{2}\right) e^{-2k}}{n^{3}},  \label{eq:FsGen}
\end{equation}
where $k=\pi nl/a$\cite{Sridhar}. The series is calculated with the result

\begin{eqnarray}
\frac{F_{stray}}{{\cal A}} &=&\frac{2}{\pi ^{3}}Mu_{0}^{2}a\Bigl[\zeta
\left( 3\right) -%
\mathop{\rm Re}%
Li_{3}\left( -e^{i\pi \delta }\right)  \nonumber \\
&&-Li_{3}\left( e^{-b}\right) +%
\mathop{\rm Re}%
Li_{3}\left( -e^{-b+i\pi \delta }\right)  \nonumber \\
&&+\frac{b^{2}}{2}%
\mathop{\rm Re}%
\ln \frac{1-e^{-b}}{1+e^{-b+i\pi \delta }}\Bigr],  \label{eq:Fstray}
\end{eqnarray}
where $Li_{n}\left( z\right) \equiv \sum_{s=1}^{\infty }z^{s}/s^{n}=z\Phi
(z,n,1)$\cite{Gradstein}, $b=2\pi l/a$. For $\delta =0$ this formula gives
the total elastic energy of a symmetric domain structure (a pattern of
opposite domains of equal widths). We have found earlier a somewhat similar
formula for another symmetric domain structure of $a_{1}|a_{2}$ type\cite
{BLxy}. Let us consider the following general cases.

{\em Zero extrinsic misfit }$\left( w=0\right) .-$ In this case the energy
of the {\em homogeneous} elastic field is simply $F_{h}/{\cal A}%
=Mu_{0}^{2}l\delta ^{2}.$ The {\em stray} energy has a simple form for the
following two limiting cases. In the standard case of narrow domains $\left(
a\ll l\right) $ and $\delta \ll 1$

\begin{equation}
\frac{F_{stray}}{{\cal A}}=\frac{2}{\pi ^{3}}Mu_{0}^{2}a\left[ \frac{7}{4}%
\zeta \left( 3\right) -\frac{\pi ^{2}\ln 2}{2}\delta ^{2}+\frac{\pi ^{4}}{96}%
\delta ^{4}\right] .  \label{eq:FstrK}
\end{equation}
One has to add the surface energy of the domain walls to find the
equilibrium domain width 
\begin{equation}
F_{dw}/{\cal A}=\sqrt{2}\gamma l/a,  \label{eq:Fdw}
\end{equation}
where $\gamma \equiv Mu_{0}^{2}\Delta /\sqrt{2}$ is the energy of the unit
surface of the domain walls, and $\Delta $ is a characteristic microscopic
length \cite{Delta}. The total free energy, $F=F_{h}+F_{stray}+F_{dw},$ is
minimal for symmetric domain structure ($\delta =0$ and $a_{1}=a_{2}=a)$,
with the standard (Kittel)\ domain width (cf.\cite{Roitburd76}) 
\begin{equation}
a=a_{K}\equiv \left( \frac{2\pi ^{3}}{7\zeta (3)}l\Delta \right) ^{1/2}\sim 
\sqrt{l\Delta }\ll l.  \label{eq:ak}
\end{equation}
Using this result, we see immediately that the high-frequency response of
the domain pattern (at a fixed domain width $a)$ is 
\begin{equation}
\kappa _{\infty }=\frac{\partial ^{2}}{\partial \delta ^{2}}\left( \frac{F}{%
{\cal A}}\right) =2Mu_{0}^{2}l\left( 1-\frac{\ln 2}{\pi }\frac{a}{l}+\frac{%
\pi a}{8l}\delta ^{2}\right) >0,  \label{eq:kappa}
\end{equation}
so the pattern is stable, although it softens because of the negative
contribution to the stray energy from the terms $\propto \delta ^{2}$ in (\ref
{eq:FstrK})$.$ Note also that $\kappa _{\infty }$ increases with $\delta ,$
and this, as we shall see, is a general result.

If the elastic modulus $\mu $ were soft, the domains can become {\em wide}, $%
a\gg l$ $\left( b\ll 1\right) $. There

\begin{equation}
\frac{F_{stray}}{{\cal A}}=Mu_{0}^{2}l\left[ 1-\delta ^{2}\left( 1-\frac{\pi
l}{a}\right) -\frac{8l}{\pi a}\ln \frac{e^{3/4}a}{4\pi l}\right] ,
\label{eq:FstrNK}
\end{equation}
and one finds a symmetric structure with the large domain width \cite{BLxy} 
\begin{equation}
a=4\pi e^{1/4}l\exp \left( \pi \Delta /8l\right) ,  \label{eq:awide}
\end{equation}
which would be $\gg l$ if we had a substrate with $\Delta \gg d_{at}$ (cf.
the answer for a ferroelectric capacitor \cite{BL1}). This is contrary to
a previously considered Kittel case with narrow domains $a_{K}\ll l.$ In the
case of the wide domains the response softens considerably, but remains
positive, 
\begin{equation}
\kappa _{\infty }=2\pi Mu_{0}^{2}l\frac{l}{a}>0.  \label{eq:kwide}
\end{equation}
It is unlikely, however, that the softness of the present domain structure
is of any practical importance. The problem is that even a small extrinsic
misfit in very thin films will push the domain structure to the boundary of
its existence \cite{Pompe,Pertsev96}, where it becomes {\it rigid} for
experimentally accessible frequencies of external field, as discussed below.

{\em \ Non-zero extrinsic misfit }$\left( w\neq 0\right) .-$ The energy of
homogeneous stresses is given by Eq. (\ref{eq:Fh}). To find the domain
structure close to the boundary of its stability we need all characteristics
in the limit $\delta \rightarrow 1,$ and $a\gg l$ ($b\ll 1),$ i.e. close to
a monodomain state. The stray energy in this limit is 
\begin{equation}
\frac{F_{stray}}{{\cal A}}=\frac{1}{4}Mu_{0}^{2}\left( 1-\delta ^{2}\right)
^{2}\left[ \ln \frac{4el}{a\left( 1-\delta ^{2}\right) }-\frac{\pi l}{a}%
\right] .  \label{eq:Fd1NK}
\end{equation}
Note that Roytburd has approximated the numerically computed stray energy
with the functional dependence $\left( 1-\delta ^{2}\right) ^{2},$ i.e.
without the important log term\cite{RoytburdYu,Roytburd98}. Minimizing the
total energy, $F_{tot}=F_{h}+F_{stray}+F_{dw},$ with respect to the
half-period of the structure $a$, we obtain the equation 
\begin{equation}
\frac{\left( 1-\delta ^{2}\right) ^{2}}{4\pi }\ln \frac{4l}{a\left( 1-\delta
^{2}\right) }=\frac{l\Delta }{a^{2}},  \label{eq:a}
\end{equation}
which has a solution 
\begin{equation}
\frac{a}{l}=\frac{1}{\left( 1-\delta ^{2}\right) \ln ^{1/2}\left( 4l/\pi
\Delta \right) }\sqrt{\frac{8\pi \Delta }{l}},  \label{eq:al}
\end{equation}
where we have omitted the terms $\propto \ln \ln \left( 4l/\pi \Delta \right)
\ll \ln \left( 4l/\pi \Delta \right) .$ It shows that the domain period
diverges as $1/(1-\delta ^{2})$ close to the phase boundary with the
monodomain state. The total energy then takes the form 
\begin{equation}
\frac{F_{tot}}{Mu_{0}^{2}l{\cal A}}=(\delta -\delta _{0})^{2}+\phi \left(
1-\delta ^{2}\right) -\frac{1}{4}\left( 1-\delta ^{2}\right) ^{2},
\label{eq:Ftotd1}
\end{equation}
where from the equilibrium value of $\delta $ is to be found (subject to the
constraint $\delta ^{2}\leq 1)$, with the parameter $\phi \equiv \left( 
{\cal L}\Delta /2\pi l\right) ^{1/2}\ll 1,$ and ${\cal L}=\ln \left( 4l/\pi
\Delta \right) $ the logarithm of a large number $\sim l/\Delta \gg 1.$

The transition into a monodomain state $\left( \delta ^{2}=1\right) $ occurs
as a function of the extrinsic relative misfit at $\left| \delta _{0}\right|
=1-\phi .$ This condition suggests a critical{\em \ }thickness of the film $%
l_{c},$ where the polydomain-monodomain transition takes place at a given
misfit $\delta _{0}$ (i.e. the {\em phase boundary} in $\delta _{0}-l$
plane), 
\begin{equation}
\frac{l_{c}}{\Delta }=\frac{1}{\pi \left( 1-\delta _{0}\right) ^{2}}\ln 
\frac{1}{1-\delta _{0}}.  \label{eq:lc}
\end{equation}
Thus, the domain structure may exist only at $l>l_{c}\left( \delta
_{0}\right) .$ This formula is obtained for $l_{c}/\Delta \gg 1$, but it
also correctly gives an {\em absence of the critical thickness} of the film
when the parameter $\delta _{0}$ is zero ($l_{c}=0$ when $\delta _{0}=0).$
The absence of the critical film thickness has been suggested earlier for
another domain pattern, $a_{1}|a_{2},$ from numerical computations\cite
{Pompe}. This is contrary to the speculations by Roytburd, who obtained $%
l_{c}\neq 0$ for $\delta _{0}=0,$ apparently as an artifact of the employed
approximations\cite{Roytburd98}. The critical point is approached linearly
with $\delta _{0}$%
\begin{equation}
\delta =1-\left( 1-\phi -\delta _{0}\right) /\phi .  \label{eq:dcrit}
\end{equation}
We see that the slope is $d\delta /d\delta _{0}=1/\phi \gg 1,$ so that the
approach to the critical point is very steep, it looks almost discontinuous.

We readily obtain the high-frequency (measurable)\ dynamical stiffness close
to the phase boundary, 
\begin{eqnarray}
\frac{\kappa _{\infty }}{Mu_{0}^{2}l} &=&\frac{1}{1-\delta ^{2}}\sqrt{\frac{%
8\Delta }{\pi {\cal L}l}}\ln \frac{2e^{2}{\cal L}l}{\pi \Delta }  \nonumber
\\
&\simeq &\frac{1}{1-\delta ^{2}}\sqrt{\frac{8{\cal L}\Delta }{\pi l}}%
\rightarrow \infty \quad (!).  \label{eq:kappacrit}
\end{eqnarray}
which {\em diverges} when $\delta ^{2}\rightarrow 1,$ i.e. close to the
phase boundary. This is in striking disagreement with the results of Pertsev 
{\it et al.}\cite{Pertsev96} who claimed a softness of the $c/a$ domain
structure close to the phase boundary, which actually does not materialize.
The static stiffness, which has no practical significance since it requires
very long time to relax the pattern in order to optimize the number of
domain walls, sharply vanishes with increasing relative misfit $\delta _{0}$%
, 
\begin{equation}
\frac{\kappa _{0}}{Mu_{0}^{2}l}\propto 1-\delta ^{2}\propto 2\left( 1-\phi
-\delta _{0}\right) /\phi \rightarrow 0,
\end{equation}
see the exact result in Fig. 2. Interestingly, close to the transition into
monodomain state the system splits into two groups of wide ($a_{1})$ and
narrow ($a_{2})$ domains with 
\begin{eqnarray}
a_{1} &=&\frac{1}{1-\delta }\sqrt{\frac{8\pi l\Delta }{{\cal L}}}\sim \frac{%
a_{K}}{(1-\delta )\sqrt{{\cal L}}}\rightarrow \infty ,  \label{eq:a1} \\
a_{2} &=&\sqrt{\frac{2\pi l\Delta }{{\cal L}}}\sim \frac{a_{K}}{\sqrt{{\cal L%
}}}<a_{K.}  \label{eq:a2}
\end{eqnarray}
%
%
\begin{figure}[t]
\epsfxsize=3.4in \epsffile{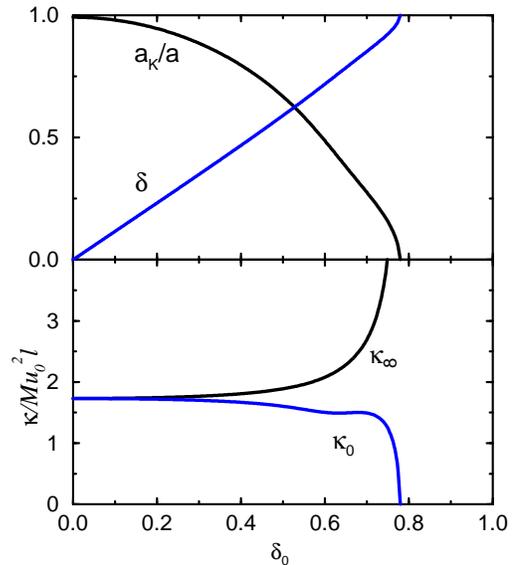}
\caption{The relative half-period of the structure $a/a_{K}$, where
$a_{K}$, Eq.(\ref{eq:ak}), 
is the standard Kittel domain width, and the asymmetry parameter $\protect%
\delta =(a_{1}-a_{2})/2a$ versus the relative extrinsic misfit $\protect%
\delta _{0}$ (top panel). Close to the boundary of existence of the domain
structure, where $a\rightarrow \infty $, the static stiffness $\protect%
\kappa _{0}$ vanishes, whereas the (measurable) dynamic stiffness $\protect%
\kappa _{\infty }$ diverges (bottom panel). }
\label{fig:fig2}
\end{figure}
We see that the width of the wide domains diverges, the density of the
domains walls decreases, but the width of the narrow domains, $a_{2},$
remains small. The narrow domains are somewhat compressed compared to the
standard Kittel width $a_{K}$ (\ref{eq:ak}).

It becomes obvious from the present analysis that it is unlikely that one
can succeed in making a ``soft'' domain pattern with the small $\kappa
_{\infty }$ in epitaxial thin ferroelectric films with elastic domains,
suitable for the applications. In fact, the only way to do this would be to
avoid any extrinsic misfit, i.e. to keep $\delta _{0}=0,$ and to reduce the
film thickness. However, in this case the interval of the extrinsic misfit
strain, which allows the very existence of the domains, is very narrow.
Then, the proximity to the phase boundary in systems with two kinds of
inequivalent domains means that the stiffness rapidly increases, oppositely
to what one actually desired. We have tacitly assumed above that the
stiffness of the domain structure determines its dielectric response.
Indeed, one can assume that there is a spontaneous polarization parallel to
the film plane in one domain and perpendicular to it in the other in the
present $c/a$ domain structure (as in $c$-domains in perovskites). One can
also assume, as in Ref. \cite{Pertsev96}, that $\delta -\delta
_{eq}=lP_{s}E_{ext}/\kappa _{\infty }$ in external field, where $P_{s}$ is
the spontaneous polarization, $E_{ext}$ the external electric field, and $%
\delta _{eq}$ is the equilibrium value of $\delta $ in zero field. We have
shown earlier \cite{BLxy} that this relation may not hold, and in some cases
it strongly overestimates the dielectric response. Additionally, the
neglected contribution of uncompensated electric fields would only make the
response stiffer. However, even this overestimate leads one to a conclusion
that the dielectric response of the pattern with inequivalent elastic
domains in the epitaxial thin films is actually suppressed.


\begin{references}
\bibitem{Delta}  $\Delta $ is usually considerably smaller than the atomic
length scale $d_{at}$ for proper ferroelastics. For perovskite
ferroelectrics, which are improper ferroelastics, it is comparable to the
domain wall width. We absorb $\sqrt{2}$, the factor coming from the domain
walls being inclined at 45$^{\circ }$, into the definition of $\Delta $.
\end{references}
\end{document}